%% file: main.tex
\documentclass[conference]{IEEEtran}
\IEEEoverridecommandlockouts
\usepackage{cite}
\usepackage{amsmath,amssymb,amsfonts}
\usepackage{algorithmic}
\usepackage{graphicx}
\graphicspath{{./img/}}
\usepackage{textcomp}
\usepackage{xcolor}
\usepackage{hyperref}

\def\code#1{\texttt{#1}}

\def\BibTeX{{\rm B\kern-.05em{\sc i\kern-.025em b}\kern-.08em
    T\kern-.1667em\lower.7ex\hbox{E}\kern-.125emX}}
\begin{document}

\title{Automatic Particle Trajectory Classification in Plasma Simulations
}

\author{\IEEEauthorblockN{Stefano Markidis$^1$, Ivy Peng$^2$, Artur Podobas$^1$, Itthinat Jongsuebchoke$^1$, Gabriel Bengtsson$^1$, and Pawel Herman$^1$}
\IEEEauthorblockA{\textit{$^1$KTH Royal Institute of Technology, Stockholm, Sweden} 
}
\IEEEauthorblockA{\textit{$^2$Lawrence Livermore National Laboratory, Livermore, CA, USA} 
}
}

\maketitle

\begin{abstract}
Numerical simulations of plasma flows are crucial for advancing our understanding of microscopic processes that drive the global plasma dynamics in fusion devices, space, and astrophysical systems. Identifying and classifying particle trajectories allows us to determine specific on-going acceleration mechanisms, shedding light on essential plasma processes.

Our overall goal is to provide a general workflow for exploring particle trajectory space and automatically classifying particle trajectories from plasma simulations in an unsupervised manner. We combine pre-processing techniques, such as Fast Fourier Transform (FFT), with Machine Learning methods, such as Principal Component Analysis (PCA), k-means clustering algorithms, and silhouette analysis. We demonstrate our workflow by classifying electron trajectories during magnetic reconnection problem. Our method successfully recovers existing results from previous literature without a priori knowledge of the underlying system.

Our workflow can be applied to analyzing particle trajectories in different phenomena, from magnetic reconnection, shocks to magnetospheric flows. The workflow has no dependence on any physics model and can identify particle trajectories and acceleration mechanisms that were not detected before.

\end{abstract}


\input{introduction}
\input{background}
\input{method}
\input{setup}
\input{explore}
\input{results}
\input{related}
\input{discussion}

\section*{Acknowledgments}
Funding for the work is received from the European Commission H2020 program, Grant Agreement No. 801039 (EPiGRAM-HS, \url{https://epigram-hs.eu/}) and Grant Agreement No. 800904 (VESTEC, \url{https://vestec-project.eu/}). LLNL-CONF-814979.

\bibliographystyle{IEEEtran}
\bibliography{main}
\end{document}

%% file: introduction.tex
\section{Introduction}
Large-scale plasma simulations are among the most important tools for understanding plasma dynamics in fusion devices, such as tokamaks, space, and astrophysical systems. Various kinds of acceleration and heating mechanisms are present in a plasma due to instabilities and interaction between waves and plasma particles (electrons and protons). Different acceleration mechanisms are associated with specific particle trajectories. Identifying and classifying the particle trajectories would allow us to understand the fundamental microscopic processes that drive the global dynamics in a plasma. 

Given an initial dataset of particle trajectories, the classification and characterization of different particle trajectories (proxies of different acceleration mechanisms) is a formidable task for at least two reasons. 

The first reason is that we do not know a priori different trajectory classes for most plasma configurations. Analytical models of particle trajectories exist for specific phenomena in simplified geometry. However, they might introduce approximations, or they might not be comprehensive. Some tools can be devised to monitor and track particles only in a very localized region of space or with a specific energy state~\cite{peng2015energetic}. However, these tools are problem-specific. They often require adaptions and new calibrations to detect particle trajectories classes in a different system configuration. Moreover, in most of the cases, particle trajectory datasets are unlabeled. For these reasons, an unsupervised method would be a convenient tool to investigate the possibility of categorizing different particle trajectories during various phenomena. 

The second challenge in particle trajectory classification is the massive amount of data that need to be analyzed. Typical plasma simulations on HPC systems use billions of particles. They can quickly generate TB-size datasets when particle trajectories are saved to disk. Manual classifications of particle trajectories are not efficient, if not impossible. An automatic procedure is essential for productive data exploration and scientific discoveries from such large-scale simulations. 

The overall goal of this work is to address these challenges by providing a workflow for identifying and categorizing particle trajectories from datasets from plasma simulations. The workflow automatically classifies particle trajectories using Machine Learning (ML) unsupervised techniques. We describe how to pre-process trajectory data, classify particle trajectories using clustering techniques, such as k-means, and select representative particle trajectories and meaningful metrics for evaluating accuracy in this work.

Our workflow is designed for data exploration of particle trajectories from Particle-in-Cell (PIC) simulations or other particle-based numerical methods \cite{hockney1988computer}. Our approach is general and does not assume a priori knowledge of the problem. We apply the workflow in this work to study electron trajectories in a specific phenomenon called magnetic reconnection. However, the same approach can also be used to study other physics processes, such as particle trajectories in shocks~\cite{peng2015kinetic}, magnetospheric flows~\cite{peng2015formation,chen2020magnetohydrodynamic}, turbulent flows~\cite{spitkovsky2008particle}, interaction of solar with lunar surface~\cite{deca2015general}, and flux ropes in tokamak devices~\cite{markidis2014signatures}. We also demonstrate that our framework can easily support the development of anomalous trajectory detection based on PCA results. By comparing the reconstruction error using a reduced number of Principle Components with a threshold value, we detect anomalous trajectories characterized by a sudden increase in the orbit oscillation frequency. 

The main contributions of this work are the following:
\begin{itemize}
    \item We develop a general workflow for automatic particle trajectory classification to categorize different kinds of particle trajectories in plasma simulations in an unsupervised manner.
    \item We integrate pre-processing (FFT and normalization) and ML techniques (Principal Component Analysis, k-means, and silhouette analysis) in our method. The workflow explores the space of particle trajectories in plasma simulations and provides characteristic trajectories.
    \item We demonstrate the workflow in one important physics phenomenon called magnetic reconnection. We validate our approach by studying the electron trajectories from a two-dimensional simulation and recovering existing classification known in the literature.
\end{itemize}

The paper is organized as follows. In Section~\ref{background}, we describe the PIC simulations that produce particle trajectory datasets. We also introduce the magnetic reconnection phenomenon. In Section~\ref{workflow}, we describe the four steps in our workflow for automatic trajectory classification. Section~\ref{setup} describes the experimental setup. Sections~\ref{explore} and~\ref{anomalous} discuss the data exploration of particle trajectories and anomaly detection with PCA. We present the classification results in Section~\ref{results} for the magnetic reconnection problem. In Section~\ref{related}, we introduce related works. Finally, we summarize our work and outline future work in Section~\ref{discussion}.

%% file: background.tex
\section{Background}
\label{background}
In this section, we introduce the simulation method to perform plasma simulations and extract particle trajectories for the classification task, and we briefly describe the use case we apply our workflow.

\begin{figure*}
    \centering
    \includegraphics[width=\linewidth]{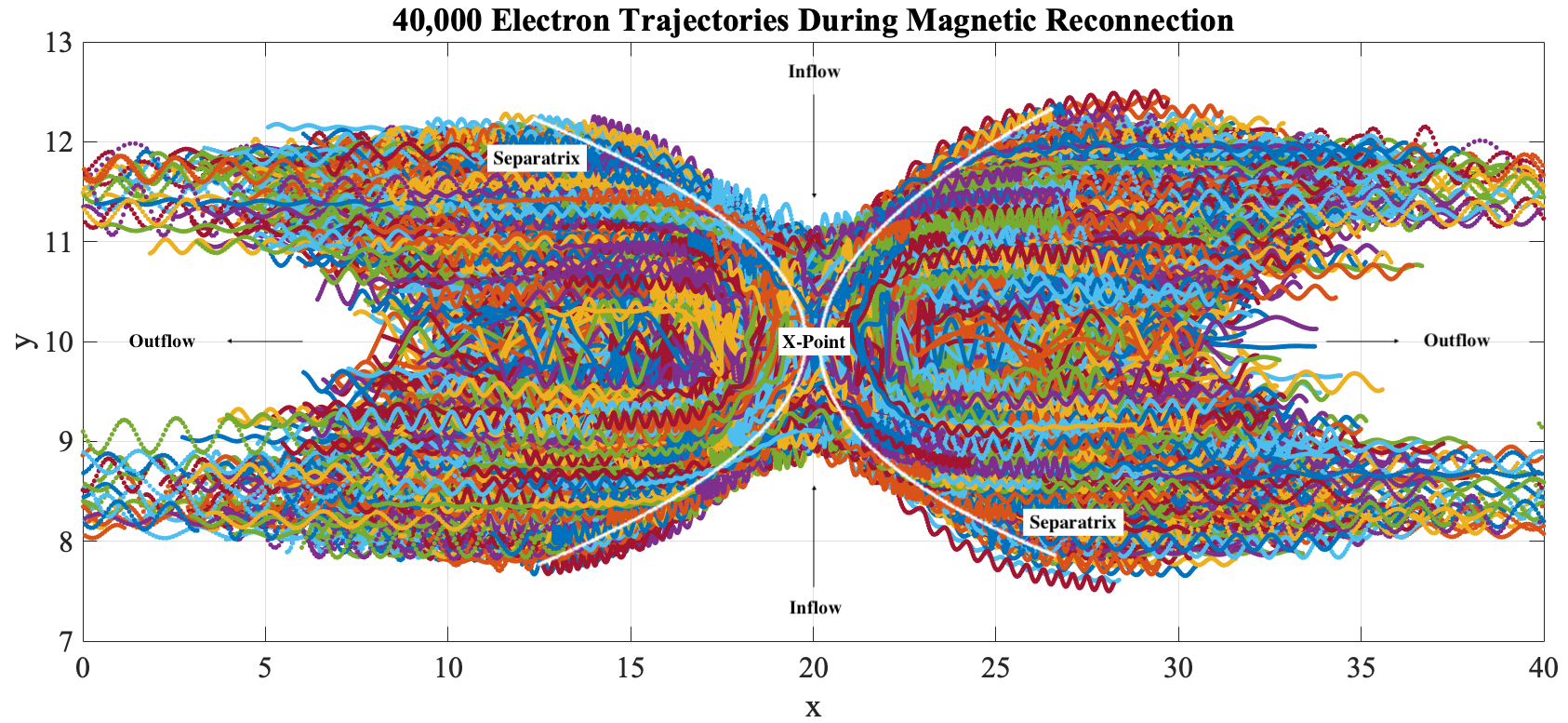}
    \caption{Electron trajectories during magnetic reconnection. During magnetic reconnection, plasma moves from the inflow regions towards the X-point and is expelled along the outflow directions. The separatrices are interfaces between two different magnetic topologies. Electron trajectories are different depending on the acceleration mechanism they undergo.}
    \label{automatic_class}
\end{figure*}

\subsection{Particle-in-Cell Simulations}
One of the most powerful and successful tools for performing plasma simulations is the PIC method. In summary, the PIC method determines the trajectories of particles (electrons and protons) under the effect of a self-consistent electromagnetic field: electrons and protons generate electric and magnetic fields that act on particles themselves. 

In this work, we use sputniPIC~\cite{chien2020sputnipic}, the successor of the iPIC3D code \cite{markidis2010multi,peng2015energetic}, to generate the particle trajectory dataset. The sputniPIC code is based on the implicit discretization of governing equations for electromagnetic kinetic equations of particles: particle equation of motion and Maxwell equations. The main computational kernel is the so-called \emph{particle mover} or \emph{pusher} that updates particle position and velocity by solving the equation of motion. Instead of the more common leap-frog or Boris particle mover~\cite{birdsall2004plasma}, we use a predictor-corrector scheme to solve the average particle velocity $\mathbf{\bar{v}}_p = (\mathbf{v}_p^{n} + \mathbf{v}_p^{n+1})/2$ during the time step $\Delta t$ with $n$ indicating the time level:
\begin{eqnarray}
\label{vhat2}
\tilde{\mathbf{v}}_p&=&\mathbf{v}_p^n+\frac{q\Delta t}{2m}\mathbf{\bar{E}}_p\\
\label{vn+1/2}
\mathbf{\bar{v}}_p&=&\frac{\tilde{\mathbf{v}}_p+\frac{q\Delta
t}{2m c}\bigl(\tilde{\mathbf{v}}_p\times\mathbf{\bar{B}}_p+\frac{q\Delta
t}{2m c}(\tilde{\mathbf{v}}_p\cdot\mathbf{\bar{B}}_p)\mathbf{\bar{B}}_p\bigr)}{(1+\frac{q^2\Delta t^2}{4m^2c^2}{\bar{B}_p}^2)},
\end{eqnarray}

where $p$ is the particle index, $q,m$ are the particle charge and mass, and $c$ is the speed of light in vacuum.
The number of iterations to determine $\mathbf{\bar{v}}_p$ is either set by a prescribed error tolerance or fixed to a small number of iterations. In this work, we use three iterations for both electron and proton particles.  The $\mathbf{\bar{v}}_p$ calculation requires the electric and magnetic field at the particle position, $\mathbf{E}_p$ and $\mathbf{B}_p$. However, the electric and magnetic field values, $\mathbf{E}_g$ and $\mathbf{B}_g$ are only defined at the grid points in the PIC method. To calculate these quantities, the PIC method uses the linear weight or interpolation functions $W({\bf x}_g-{\bf x}_p)$ defined as follows:
\begin{equation}
\label{interpW}
W({\bf x}_g-{\bf x}_p) =
\left\{
\begin{array}{l}
 1 - |{\bf x}_g-{\bf x}_p|/\Delta x \quad \textup{if} \quad |{\bf x}_g-{\bf x}_p| < \Delta x \\
0  \quad \textup{otherwise}  .\end{array}
\right.
\end{equation}
With the usage of interpolation functions, we can calculate the electric and magnetic field at the particle position from these values on the grid point $g$:
\begin{equation}
{\bf E}_p=\sum_g^{N_g} {\bf E}_g W({\bf x}_g-{\bf x}_p) \quad \quad {\bf B}_p=\sum_g^{N_g} {\bf B}_g W({\bf x}_g-{\bf x}_p) .
\label{interp}
\end{equation}
Once the particle average velocity is calculated, each particle position and velocity is updated as follows:
\begin{equation}
\label{dif_eom2}
\left\{
\begin{array}{l}
\mathbf{v}_p^{n+1} = 2 \mathbf{\bar{v}}_p - \mathbf{v}_p^{n}  \\
\mathbf{x}_p^{n+1} = \mathbf{x}_p^{n} + \mathbf{\bar{v}}_p\Delta t.
\end{array} 
\right. 
\end{equation}
Detailed descriptions of mathematical derivation of the implicit discretized equations can be found in~\cite{markidis2014fluid,markidis2011energy}. The typical PIC simulations run on supercomputers, possibly with accelerators, and uses millions of particles.

\subsection{Magnetic Reconnection}
The proposed method is generally applicable to analyzing and classifying particle (electron or proton) trajectories in any particle-based simulations of magnetized plasmas. We apply our workflow to study electron orbits in magnetic reconnection for demonstration purposes without losing generality. 

Magnetic reconnection is a common phenomenon occurring in space, astrophysical, and fusion plasmas. Magnetic reconnection takes its name from the fact that it connects two regions with initial distinct magnetic topologies. This reconfiguration of magnetic field topology is accompanied by a conversion of magnetic field energy into high-speed jets' kinetic energy. Magnetic reconnection is responsible for driving the dynamics of Earth magnetosphere, e.g., generating aurora. It could also cause disruptions in magnetic fusion devices and limit the confinement of plasmas in tokamaks.    

The dynamics and occurrence of magnetic reconnection in nature have been extensively studied because of its importance and impact. In particular, the mechanism that converts magnetic and electric energies into kinetic particle energy, which ultimately accelerates or heats particles, is a fundamental research topic. Different acceleration mechanisms lead to different characteristic trajectories. Accordingly, the classification of particle trajectories would allow us to identify acceleration mechanisms present during magnetic reconnection.

We investigate the possibility of automatic classification by focusing on a simplified two-dimensional system configuration. The computational plasma physics community has proposed this set-up under the name of \textit{GEM challenge}~\cite{birn2001geospace}. In particular, we study the electron trajectories after magnetic reconnection has fully developed.

Figure~\ref{automatic_class} shows the superposition of 40,000 electron trajectories selected from the output of sputniPIC~\cite{chien2020sputnipic} simulations. We use this output dataset for the classification in this study. We select the electrons to be tracked by randomly picking electrons located in a box enclosing the reconnection point (also known as X-point) at different simulation time steps. X-point is where magnetic reconnection is initiated. We are interested in studying electrons accelerated during magnetic reconnection, and thus, we investigate particles close to the reconnection X-point.

At the macroscopic level, the plasma bulk flow moves from the inflow regions towards the X-point. Plasma is accelerated and diverted along the outflow direction ($y$ direction in Figure~\ref{automatic_class}), forming the so-called \textit{reconnection jets}. However, at the microscopic level, electron trajectories are highly diverse as electrons undergo different acceleration mechanisms. As highlighted in~\cite{lapenta2015separatrices}, the separatrices, the thin interface dividing the inflow and outflow plasmas, are a crucial area where acceleration and plasma dynamics take place. 

Since the Sixties and seminal work by Speiser~\cite{speiser1965particle}, scientists have been studying particle trajectories in magnetic reconnection. Up to date, the work by Zenitani and Nagai~\cite{zenitani2016particle} are among the most comprehensive studies on classifying electron trajectories during magnetic reconnection. They divide electrons trajectories into several trajectories categories depending on whether they cross the midline (in the $y$ direction in Figure~\ref{automatic_class}) or follow orbits identified by previous studies, such as Speiser's and Egedal's works~\cite{egedal2005situ}.

%% file: method.tex
\section{Automatic Particle Classification Workflow}
\label{workflow}
\begin{figure}
    \centering
    \includegraphics[width=\columnwidth]{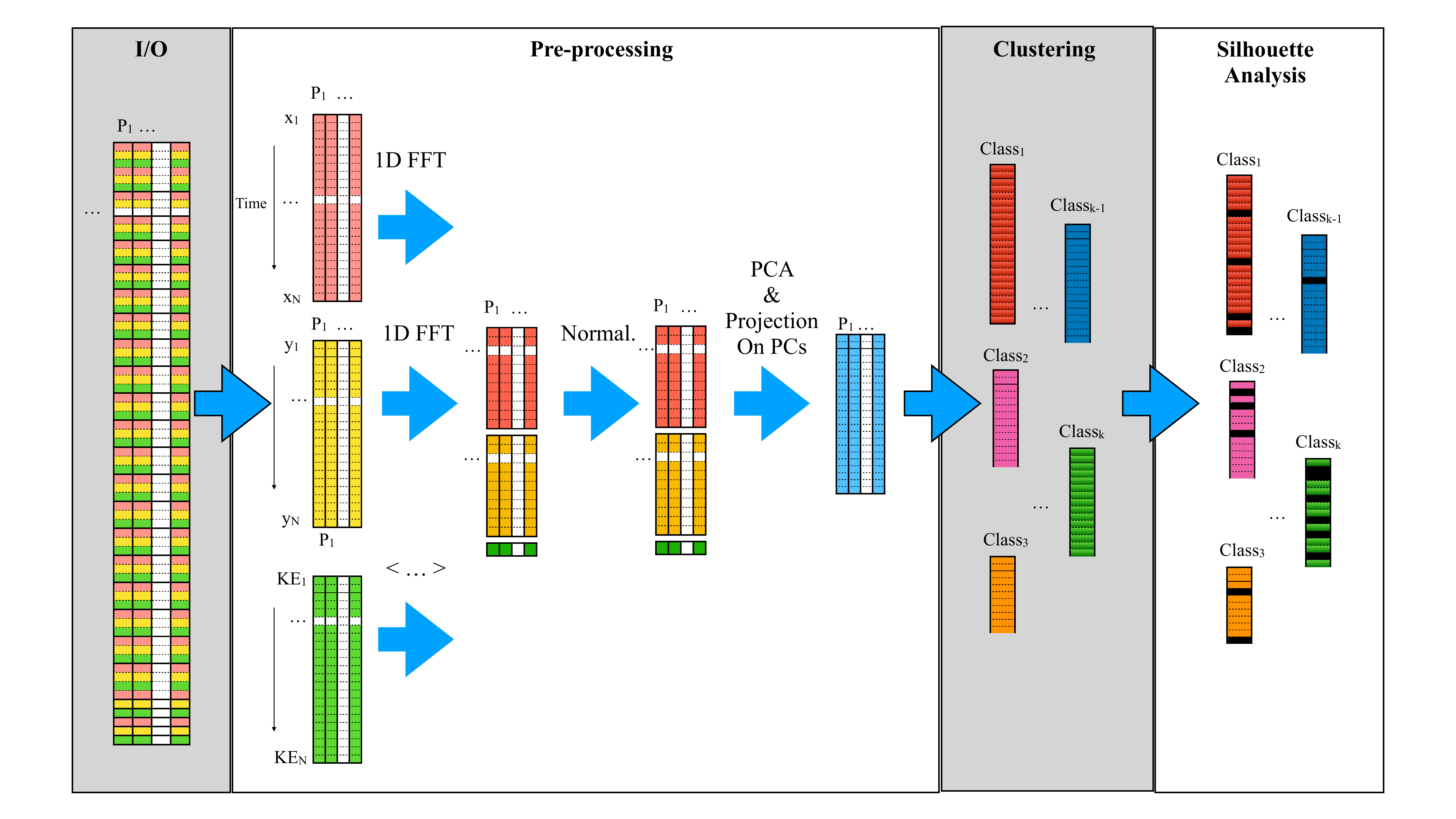}
    \caption{Our methodology consists of four steps. First, we read and extract the data of particle trajectories and associated kinetic energy, produced by sputniPIC. Second, we pre-process the data. Third, we apply k-means for clustering data in different classes. Fourth, we assess the quality of clustering by using silhouette analysis.}
    \label{workflowFig}
\end{figure}

In this study, we design and develop a workflow for classifying particle trajectories in unsupervised manner. The workflow is divided in four main phases, as shown in Figure \ref{workflowFig}.

The first step is to access particle trajectory information. Particle orbits can be either analyzed at runtime by examining the online history of particle positions in the main memory, or, as a part of the post-processing of the simulation results saved as files. In this work, we perform the analysis as part of the post-processing of our simulation. We complete several PIC simulations of magnetic reconnection and record the particle positions ($x$ and $y$ coordinates) and kinetic energies during magnetic reconnection for a total of 40,000 particles for 300 time steps.

The second phase, the \textit{pre-processing} step, focuses on preparing the data for the clustering. We divide the original dataset into three matrices. Each column of the matrix represents the $x$, $y$ coordinates, and kinetic energy for each particle. Each row includes these quantities at different time steps. For instance, for 40,000 particles trajectories recorded in 300 time steps, the three matrices have size $300 \times 40,000$. The most important stage in the pre-processing is to use 1D FFT on the particle $x$ and $y$ coordinates and express the trajectories in the spectral space. The Fourier transformation removes the spatial dependency of data, e.g., a clustering algorithm directly on the particle positions and velocities would categorize trajectories mainly depending on particle location~\cite{nyman2020exploring}. For the history of kinetic energy, we take the average to reduce the problem's dimensionality. We found that taking the FFT of the kinetic energy does not lead to any improvement in the clustering quality. After this step, we normalize the FFT results to have all the data in the 0-1 range. Then, we apply the Principal Component Analysis~(PCA) to study if there are any low-dimensional manifolds facilitating more parsimonious data representations~\cite{james2013introduction}. This last step reduces the dataset's dimensionality while still retaining almost all the dataset variance. In particular, we use 20 principal components~(PCs) that account for 98.8\% of the pre-processed data variance. We remove the spatial dependency through pre-processing, relying on the spectral representation of the dataset and reducing the problem's dimensionality from 900 ($x$,$y$ coordinates, and kinetic energy for 300 steps) to 20 (the number of the PCs).

The third phase of the workflow is the unsupervised classification using a clustering technique applied to the projection of the pre-processed data on the PCs (20 coefficients represent each trajectory). Different clustering techniques exist, e.g., Gaussian-mixture, affinity propagation, and k-means~\cite{james2013introduction}. We experimented with all these techniques. In practice, we found that k-means with cosine, city-block, and correlation distance metrics, are the most effective clustering techniques when comparing the clustering results with the trajectories classes found in the literature.

The last step is to determine how well a particle trajectory represents an identified cluster and assess the clustering quality. For this, we use the silhouette analysis \cite{kassambara2017practical} that associates a coefficient ranging from -1 to 1 to each trajectory. If the silhouette coefficient is positive and close to one, then the trajectory is highly representative of the class, while a negative coefficient represents a trajectory that might be misclassified. 

%% file: setup.tex
\section{Experimental Set-up}
\label{setup}
We use the sputniPIC PIC code for simulations to obtain the electron trajectories. We choose a well-known simulation set-up in space physics -- the GEM challenge~\cite{birn2001geospace}  -- for simulating the magnetic reconnection phenomenon in a simplified, yet realistic configuration. The simulation parameters are derived from observations of the Earth magnetotail. Our magnetic reconnection simulation uses electrons with a higher charge-to-mass ratio, 64, instead of the default 25 in the GEM challenge. The simulation box is $40 d_i \times 20 d_i$, where $d_i$ is the ion skin depth, a characteristic length at which the electron and proton dynamics decouple. The grid consists of $256 \times 128$ cells. We use four particle species: the current layer and background proton and electron populations. Each particle species is initialized with 125 particles per cell. The total number of particles is approximately $2.6E8$. For the performance evaluation, we advance the simulation for 30,000 time steps. Each time step is equal to $\omega_{pi} \Delta t = 0.25$, where $\omega_{pi}$ is the ion plasma frequency.

We perform several simulations of magnetic reconnection and save the position and kinetic energy for 40,000 electrons. The electron coordinates and kinetic energies are recorded for 300 time steps during magnetic reconnection (after 2,200, 2,400, 2,600, and 2,800 steps). The data sets of the saved electron trajectories are in \code{.csv} files \footnote{The Matlab script and the dataset are available at \url{https://github.com/smarkidis/trajectory_classification_plasma}}.

To enable the analysis and classification of electron trajectories, we use the Matlab R2020b framework. For carrying out PCA, we run the Matlab \code{pca()} function and retain the first 20 PCs that account for 98.82\% of the variance. For clustering, we rely on the Matlab \code{kmeans()} function with 50 replicates (the number of times to repeat the clustering, each with a new set of initial centroids) and 1,000 maximum number of iterations. We tested different distance metrics. The cosine, city-block and correlation distance metrics are found to provide clusters of trajectories best-reflecting trajectories known in the literature~\cite{zenitani2016particle}. To perform the silhouette analysis, we use the Matlab \code{silhouette()} function. We present the particle orbits most representative of the clusters found, i.e., with the highest silhouette coefficients.

%% file: explore.tex
\section{On the possibility of Classifying Particle Trajectories}
\label{explore}
\begin{figure}
    \centering
    \includegraphics[width=\columnwidth]{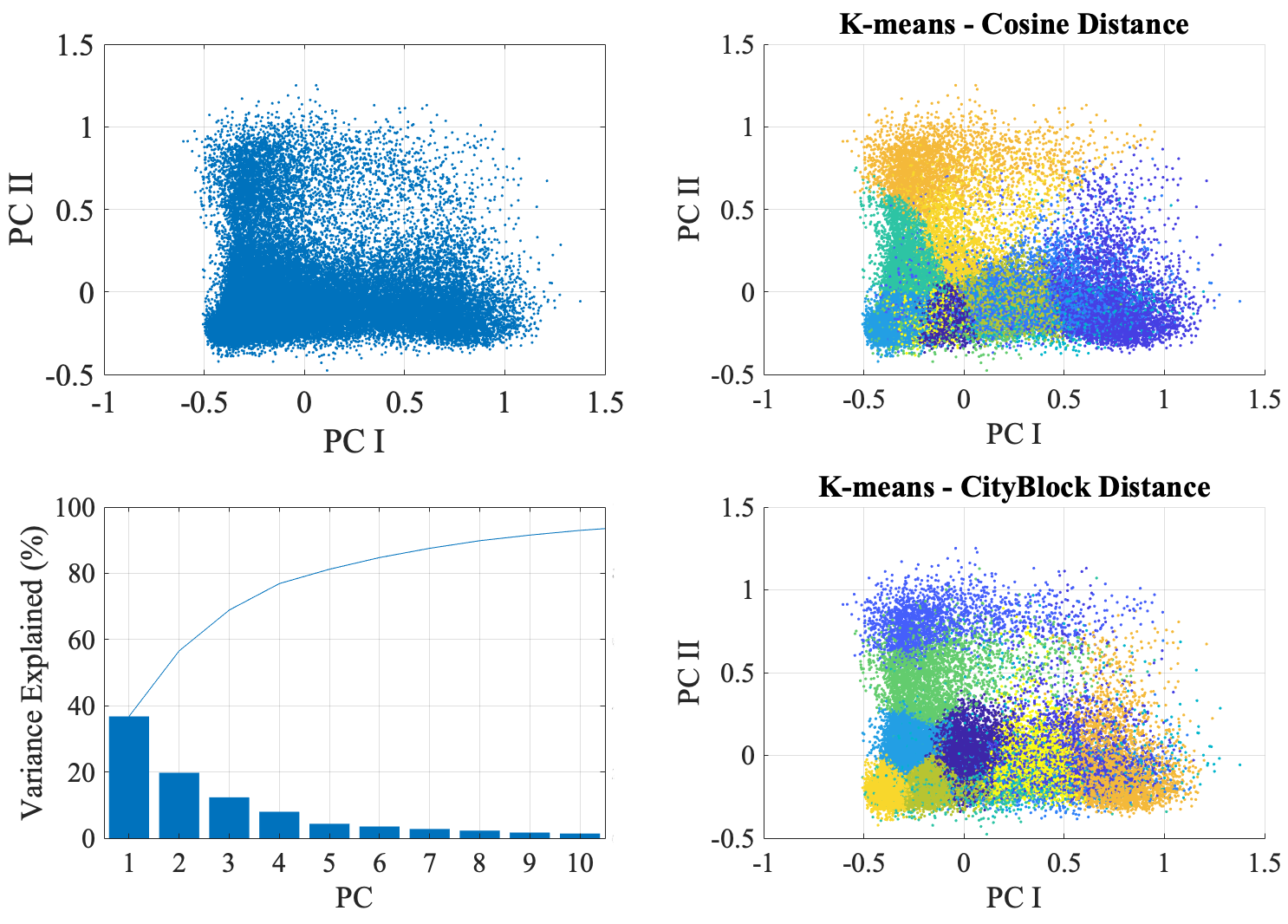}
    \caption{We perform a PCA on the pre-processed data and show the projection on the first two PCs and the scree plot on the left panels. We show how the k-means using two different distance metrics leads to two different clusterings on the right panels.}
    \label{fig:pca}
\end{figure}

Our investigation's first step is to understand whether our dataset, comprising 40,000 pre-processed electron trajectories, naturally presents structures or clusters in a low-dimensional space. For visualizing our data set in a low-dimensionality space, we plot in Figure~\ref{fig:pca} the projection of pre-processed trajectories on the first two PCs obtained by the PCA. The first two PCs account for 56.57\% of the pre-processed dataset total variance (see the scree plot for the bottom left panel in Figure~\ref{fig:pca}).

In Figure~\ref{fig:pca}, each trajectory is represented by a blue point. It is clear that there is no cluster or structure emerging in the low-dimensional space by investigating this plot. Instead, except for a few outliers, the trajectory projections are continuous in the low-dimensional space. By applying other dimensionality reduction to the pre-processed data, such as \textit{T-SNE}, we also obtained similar results, e.g., we do not observe any clear cluster. We expect this result as multiple acceleration mechanisms might be present, leading to trajectories that mix different characteristic orbits.

While it is not possible to identify by inspection the clusters in the plot using the projection on the first two PCs, we can rely on unsupervised clustering methods, such as k-means, to partition our pre-processed dataset in a given number of clusters. The results of clustering strongly depend on the distance used by the clustering methods. In fact, the use of a given distance implies space geometry, e.g., Euclidean, that does not map to the actual geometry of highly-dimensional space. The right panels of Figure~\ref{fig:pca} shows how k-means with two different distances, \textit{cosine} and \textit{city-block}, partition our pre-process dataset in 12 classes. 

Because each cluster has no clear separation from other clusters, it is crucial to identify the clusters' most representative particle trajectories and neglect particle trajectories that are a mix of different clusters. To determine the most representative particle trajectories, we use the silhouette analysis.

Clustering techniques, such as k-means, require to set the number of clusters to divide the dataset. In general, the number of trajectory classes is not a priori known, and the elbow method~\cite{james2013introduction} might fail to provide the most convenient number of classes given the fact there is a continuous transition between different particle trajectories. Our approach gives us a domain-specific knowledge to identify the correct number of categories. We start from a relatively high number of classes, e.g., 36 or 25, corresponding to distinct trajectories. With such a high number of classes, many trajectories classes are similar, and then the number of clusters can decrease. We stop reducing the number of clusters when one of the characteristic classes (with no other similar class) merges into another class. In our particular use-case of magnetic reconnection, we start with 25 classes and then reduce to 12. If we use 11 categories, one important trajectory class is merged into another category by the clustering method.

\begin{figure}[bt]
    \centering
    \includegraphics[width=\columnwidth]{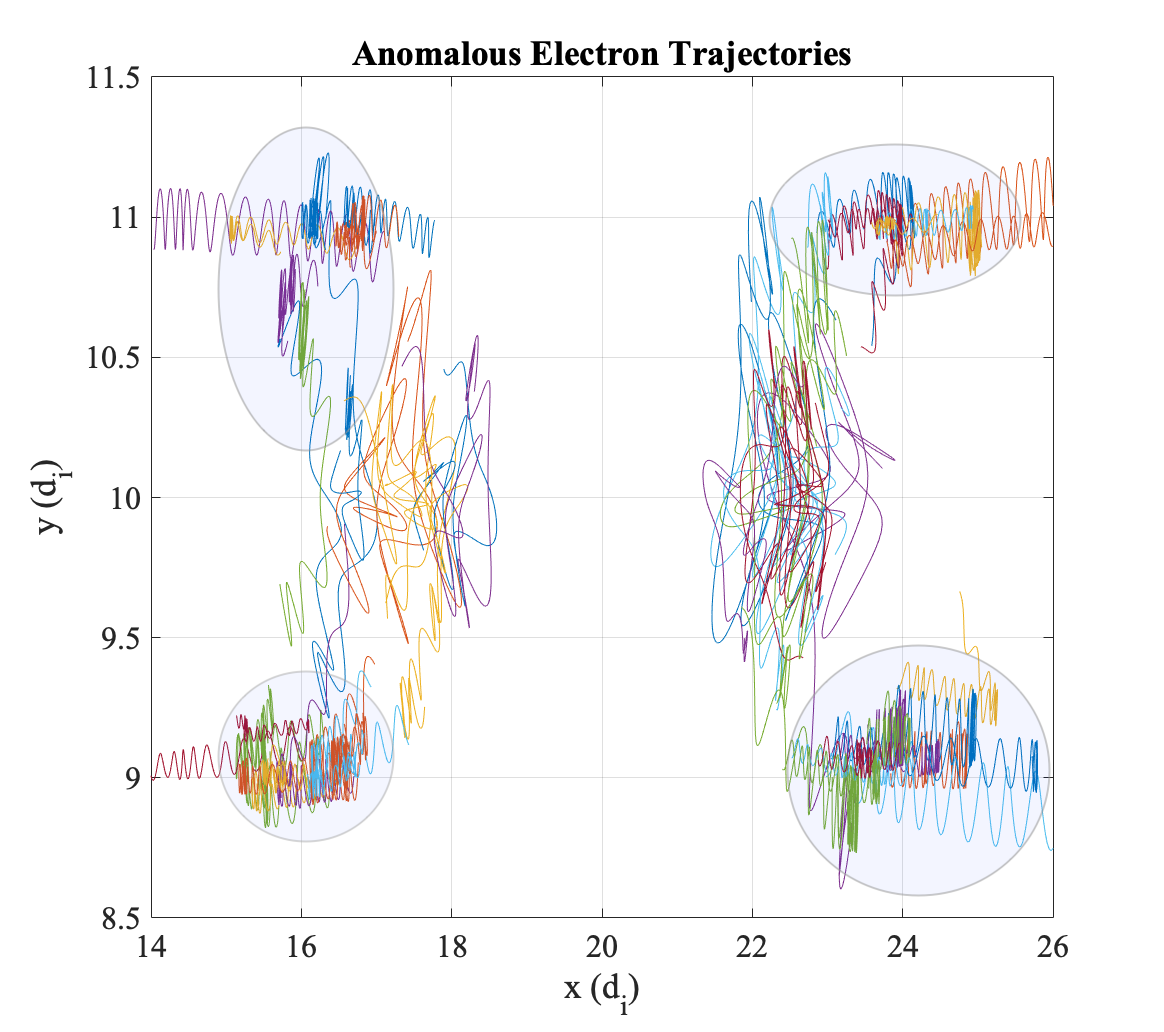}
    \caption{Electron trajectories identified as \textit{anomalous} by our PCA-based anomaly detection algorithm. The anomalous trajectories are characterized by a sudden increase in orbit oscillation frequency (see regions enclosed in the blue ellipses) that cannot be fully captured by the FFT pre-processing stage and the reduced number of PCs. This kind of orbits is often found in the proximity of shocks.}
    \label{fig:anomalous}
\end{figure}
\section{Detection of Anomalous Electron Trajectories}
\label{anomalous}
Our workflow can facilitate the design and implementation of a simple method for detecting anomalous trajectories based on PCA results. The general idea is to reconstruct the original post-processed orbits using a reduced number of PCs, e.g., 20, and calculate the reconstruction error. If the error is larger than a threshold value, we can classify the trajectories as \textit{anomalous}. For instance, we identify an orbit as anomalous if the error calculated with the Euclidean norm is higher than 0.1. Figure~\ref{fig:anomalous} shows the trajectories that suffer from a large reconstruction error when using PCA with 20 PCs. We note that all the particle trajectories, detected as \textit{anomalous}, are characterized by a sudden increase in the frequency of orbit oscillation and a consequent bouncing motion (see the regions enclosed in the blue ellipses in Figure~\ref{fig:anomalous}). This kind of orbits is often found in the proximity of shocks. High-frequency orbit oscillations are not fully captured by a finite number of Fourier modes and PCs leading to a large error in the reconstruction.  

%% file: results.tex
\begin{figure*}[ht]
    \centering
    \includegraphics[width=\linewidth]{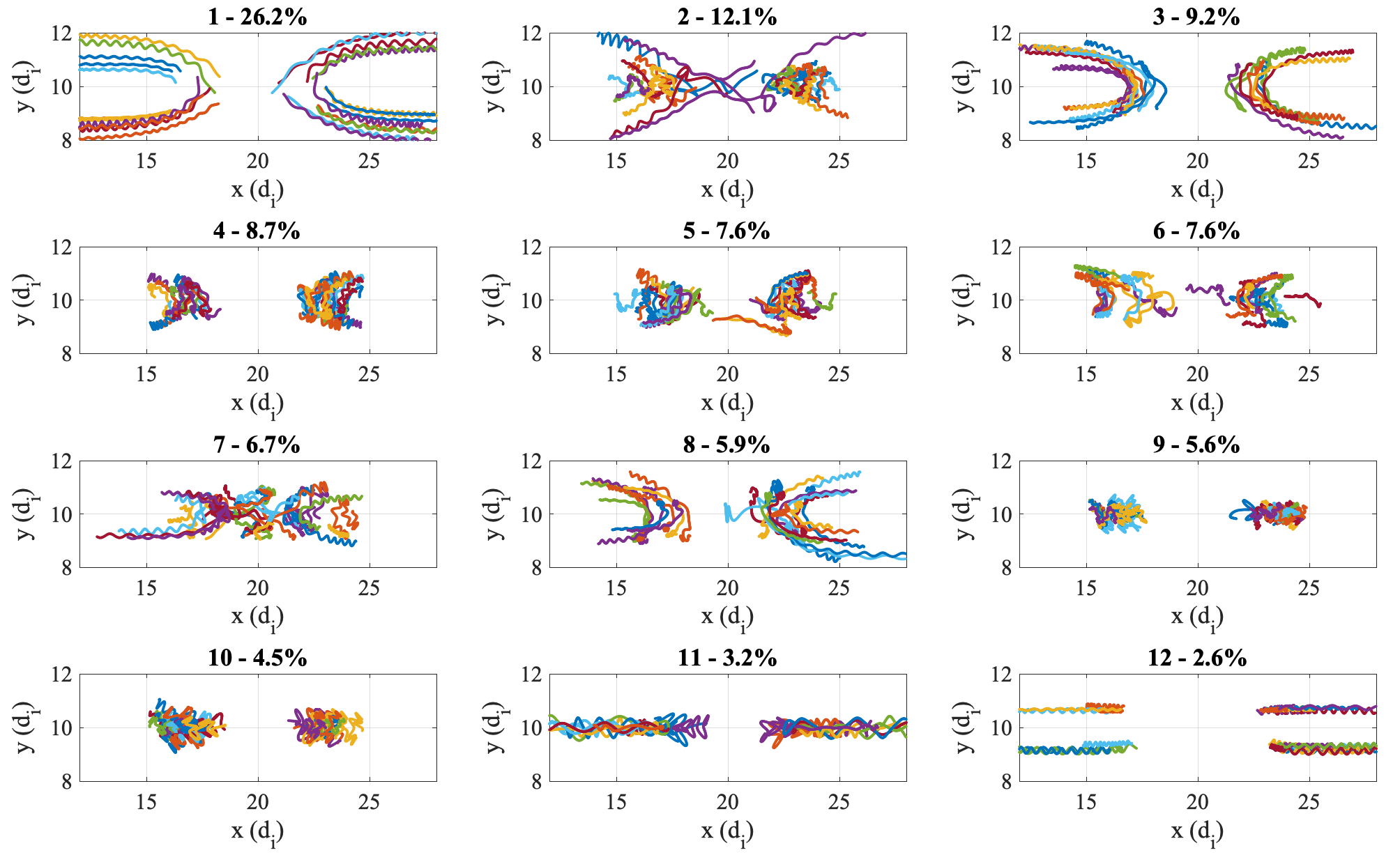}
    \caption{12 trajectories classes found by applying k-means on the trajectories using the cosine distance metric. In each panel, we show the 25 particle trajectories with the highest silhouette factor. The percentage in the title is the percentage of the orbits belonging to that class.}
    \label{automatic_class1}
\end{figure*}
\section{Classifying Electron Trajectories in Magnetic Reconnection}
\label{results}
We apply the k-means clustering method using the cosine distance metric and 12 classes to the pre-processed (FFT and PCA) dataset. Figure~\ref{automatic_class1} reports the classification results. Each panel shows the 25 trajectories with the highest silhouette score.

The clustering process divides all the trajectories into 12 classes, and each class could have a different number of particle trajectories. The title in the subplots of Figure~\ref{automatic_class1} indicates the number of the class and the percentage of trajectories belonging to the class. The percentage also includes those trajectories with low or negative silhouette score. For instance, we note that class 1 includes the electron trajectories flowing along the separatrices, and they account for 26.2\% of the total 40,000 electron trajectories. Classes 2 and 3 also include trajectories along the separatrices. Almost 50\% of electron trajectories in our dataset are located along the separatrices, showing that separatrices are the critical regions for magnetic reconnection~\cite{lapenta2015separatrices}. 

Categories 4-6 include different particle trajectories localized at the tips of the separatrices and next to the X-point. These two regions are also known as \textit{magnetic reconnection jet fronts} and feature characteristic acceleration and heating mechanisms~\cite{khotyaintsev2017energy}. 

Classes 9-11 comprise electron trajectories in the outflow region. These orbits have been identified by previous studies as \textit{nongyrotropic electrons}, \textit{local} and \textit{global Speiser orbits}. Class 12 consists of the so-called \textit{regular non-crossing} orbits. We note that when we choose 11 classes for the k-means, the category of \textit{regular non-crossing} orbits is not detected by our classifier. For this reason, we choose 12 classes for this study.

The most important result of our classification workflow is that we automatically detect characteristic electron orbits that were identified in the previous studies using an analytical or physics-based models~\cite{zenitani2016particle,speiser1965particle}. 

In our work, we also experimented k-means with city-block, correlation, and Euclidean distance metrics to assess the impact on the quality of clustering. Except for the Euclidean distance metric, we found that the usage of cosine, city-block, and correlation distances provides similar classification results that agree with trajectory classification existing in the literature. 

\begin{figure*}[ht]
    \centering
    \includegraphics[width=\linewidth]{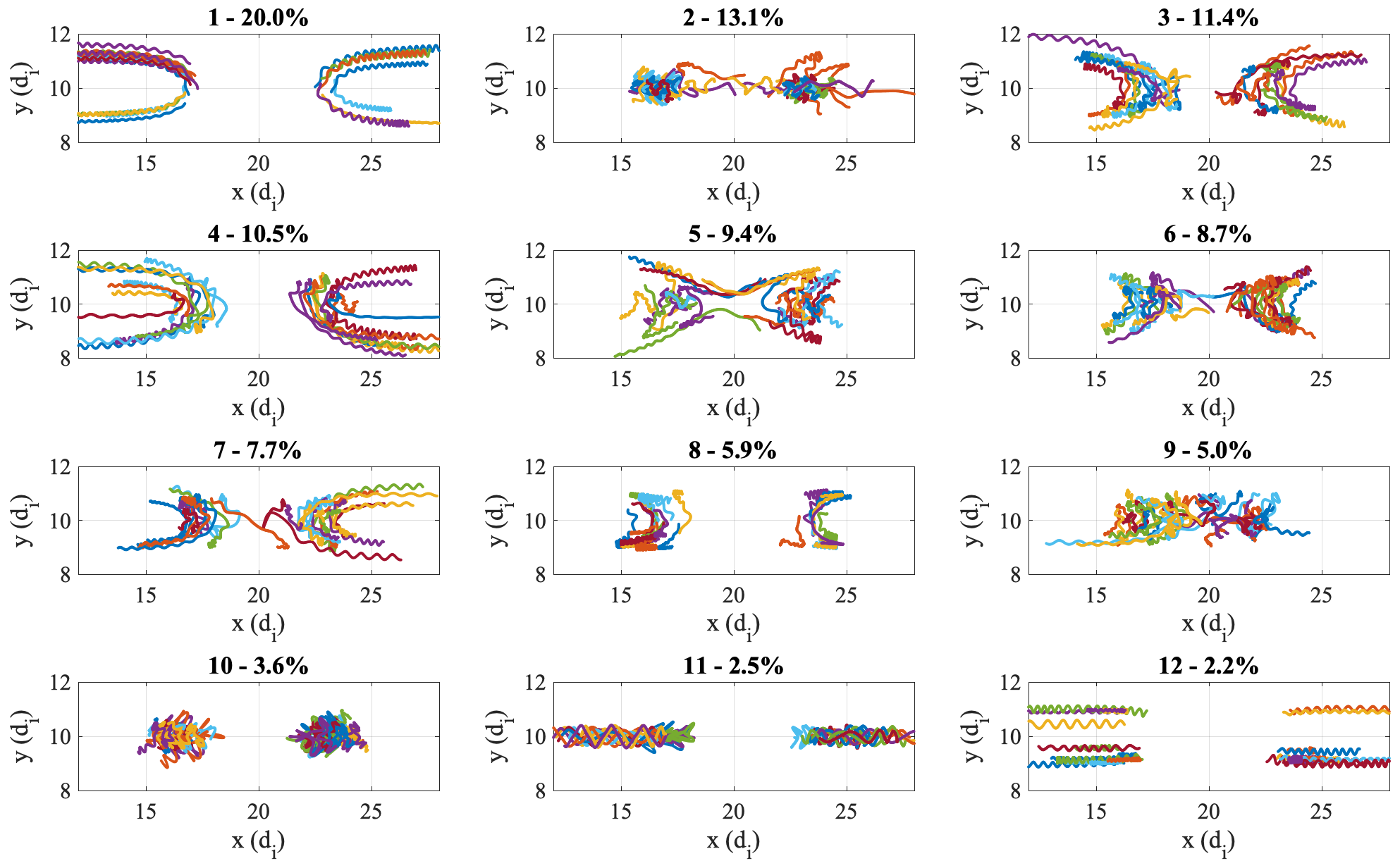}
    \caption{12 trajectories classes found by applying k-means with cityblock distance metric. The 12 categories are similar to the ones found with k-means and cosine distance. However, the percentage of trajectories belonging to a class is rather different from the percentages obtained with k-means and cosine distance metric.}
    \label{automatic_class2}
\end{figure*}
Figure~\ref{automatic_class2} shows the results of k-means clustering with the city-block distance metric. When comparing Figures~\ref{automatic_class1} and~\ref{automatic_class2}, we note that a significant difference is the percentage of particle trajectories that belong to specific classes. For instance, the first trajectory class, including particles flowing along the separatrices, accounts for 20\% of the total particle trajectories against 26.2\% as in the cosine distance case. Another difference is the larger number of particle trajectories moving between two separatrices in Class~5. 

%% file: related.tex
\section{Related Work}
\label{related}
The usage of ML techniques for classifying particles is an emerging research topic encompassing different disciplines.~\cite{angelopoulos2020particle} developed three ML models, including random tree, multi-layer perceptron model, and a convolutional neural network to detect and predict particle trajectory in nuclear physics experiments.~\cite{munoz2020single} developed a random forest architecture to associate single trajectories to the underlying diffusion mechanism characterization of experimental trajectories in biophysics experiments. Differently from these works, we focus on unsupervised learning that does not require a labeled dataset. 

\cite{patwary2016panda} introduces a framework for performing k-d tree particle classification in plasma simulations with the VPIC code~\cite{bowers20080}. Their classification is based on particle location, velocity, and energy. Instead, particle trajectory requires temporal information from a series of time steps. The temporal information regarding particle orbit complicates the classification that only considers the location of particles in one time step.

The work by Zenitani and Nagai~\cite{zenitani2016particle} provides a very comprehensive overview of different electron trajectories during magnetic reconnection. We use the results of this work to compare and validate the results of our automatic classification. In our work, trajectories classes are automatically defined and do not use any physics-based approach to classification.

%% file: discussion.tex
\section{Discussion and Conclusions}
\label{discussion}
In this work, we proposed a general workflow for classifying automatically particle trajectories from large-scale plasma simulations. The workflow performs automatic particle trajectory classification using an unsupervised approach. The significant advantage of using an unsupervised method is that it does not require a priori knowledge of existing trajectories classes or physical mechanisms in place. 

The workflow integrates four different steps, including I/O, data pre-processing using FFT and PCA, clustering using k-means, and silhouette analysis. The workflow streamlines the task from a simulation output (particle trajectories from PIC simulations) to the knowledge discovery of particle trajectories associated with characteristic acceleration and heating mechanisms in plasmas.

The crucial step in the workflow is the FFT's use on the particle trajectory data in the pre-processing stage. The FFT allows for removing spatial locality information and expressing the particle trajectories as different Fourier modes. Electron and proton motion is characterized by typical periodic motion introduced by the Lorentz force or local plasma non-neutrality (plasma oscillations). Using the FFT information, we can retain important information about particle trajectories' periodic dynamics. The proposed workflow is convenient to investigate particle orbits in plasma and other physical systems with characteristic oscillations.

We explored several pre-processing techniques applied to the original dataset containing the particle trajectories and kinetic energy. One possibility is to disregard the spatial information ($x$ and $y$ coordinates) and rely only on the kinetic energy values. Another option is to use symmetries in the system: in the 2D magnetic reconnection problem, a central symmetry with respect to the X-point exists. In this case, it is possible to mirror particle trajectories. However, we did not find an improved quality of clustering with these two pre-processing techniques when comparing the results with existing categories reported in the literature.

This work's natural next step is to reconstruct the distribution functions from the distinct trajectories that our workflow automatically identifies. The distribution function is a quantity related to the probability of finding a particle at a given position and velocity and is measured by lab experiments and spacecraft probes. An extension of our workflow to reconstruct distribution functions from particle orbits could enable a direct comparison with the distribution functions detected by lab measurements and spacecraft instruments. 
 
 We used the workflow to investigate electron trajectories during magnetic reconnection in a 2D GEM challenge simulation and successfully recovered existing known categories in literature. However, the workflow is generally applicable to explore particle trajectories in the results of any particle-based plasma simulations in a different configuration, e.g., a three-dimensional set-up of magnetic reconnection and other phenomena, e.g., shocks, turbulence, magnetospheric flows, and flux ropes dynamics. Our automatic classification workflow allows us to identify different trajectory categories, potentially unknown in the literature, and advance our understanding of the microscopic dynamics, acceleration, and heating mechanisms in plasma systems.